\documentclass[journal,12pt,draftclsnofoot,onecolumn]{IEEEtran}

\usepackage{cite}

\usepackage{makeidx,epsfig}
\usepackage{setspace,graphicx,subfigure,multirow}
\usepackage{algorithmic}

\usepackage{amsfonts,latexsym,amssymb,amsthm}
\usepackage{graphicx}
\usepackage{array}

\usepackage[cmex10]{amsmath}

\usepackage{url}

\begin{document}
\bibliographystyle{IEEEtran}

\title{Aloha Games with Spatial Reuse}
\author{Jiangbin~Lyu,
        Yong~Huat~Chew,~\IEEEmembership{Member,~IEEE}
        and~Wai-Choong~Wong,~\IEEEmembership{Senior~Member,~IEEE}%
\thanks{J. Lyu is with NUS Graduate School for Integrative Sciences and Engineering, and affiliated to the Ambient Intelligence Lab in the Interactive and Digital Media Institute (IDMI), National University of Singapore (email: jiangbin.lu@nus.edu.sg)}%
\thanks{Y. H. Chew is with the Institute for Infocomm Research, Singapore (email: chewyh@i2r.a-star.edu.sg)}%
\thanks{W. C. Wong is with the Department of Electrical and Computer Engineering, National University of Singapore (email: elewwcl@nus.edu.sg)}
}

\maketitle

\begin{abstract}
Aloha games study the transmission probabilities of a group of non-cooperative users which share a channel to transmit via the slotted Aloha protocol. This
paper extends the Aloha games to spatial reuse scenarios, and studies the system equilibrium and performance. Specifically, fixed point theory and order
theory are used to prove the existence of a least fixed point as the unique Nash equilibrium (NE) of the game and the optimal choice of all players. The
Krasovskii's method is used to construct a Lyapunov function and obtain the conditions to examine the stability of the NE. Simulations show that the
theories derived are applicable to large-scale distributed systems of complicated network topologies. An empirical relationship between the network
connectivity and the achievable total throughput is finally obtained through simulations.
\end{abstract}


\section{Introduction}
\IEEEPARstart{G}{ame} theoretic approaches have been widely used to design multiple access protocols in wireless networks. In \cite{survey}, the authors
provide a comprehensive review of the game models developed for different multiple access schemes. In particular, several channel access games in
ALOHA-like protocols are presented. For example, in \cite{MacKenzie1}\cite{ MacKenzie2}, MacKenzie and Wicker consider the slotted Aloha protocol as a game
between users contending for a conventional collision channel where no two or more users are allowed to transmit simultaneously. In their work, an infinite
users' model is adopted with a finite packet arrival rate, and all users are assumed to be indistinguishable. A strategy in this game is a mapping from the
number of backlogged users (assumed to be known to all users) to a transmission probability. The authors conclude that, for the optimal value of the cost
parameter, the throughput of a slotted-Aloha system with non-cooperative users can be as high as the throughput of a centrally controlled system. This
result is generalized in \cite{MacKenzieMPR} to show that the same result holds for multi-packet reception (MPR) channels that allow more than one packet
to be successfully received simultaneously.

An alternative Aloha game model is proposed by Jin and Kesidis \cite{alohagames}, whereby a group of heterogeneous users share a conventional collision
channel and transmit via slotted Aloha. Each user in this game attempts to obtain a target rate by updating its transmission probability in response to
observed activities. The authors further assume in \cite{alohaprice} that, for users with inelastic bandwidth requirements, each user's target rate depends
on its utility function and its willingness to pay, and they propose a pricing strategy to control the behavior of the users (in order to bring their
target rates within the feasible region). This Aloha game model is further investigated in \cite{altruistic,generic,channel}. In \cite{altruistic}, the
authors investigate the effects of altruistic behavior on the stability of equilibrium points in a two-player game. In \cite{generic}, the authors
generalize the model and propose a generic networking game with applications to circuit-switched networks. In \cite{channel}, Menache and Shimkin extend
the model by incorporating time-varying to the channel model.

The existence and stability of the equilibrium solutions have been well studied in these works. However, the results of these studies are more suitably applied to the uplink random access channel of a centralized system. There also remain fundamental issues which are unaddressed. For example, among all the equilibrium solutions, does there exist an equilibrium point which is optimal to all players, or a solution which always favors different subgroups of players?
Furthermore, if a global optimal solution does exist for all players, 
how to converge to that equilibrium point during implementation?

On the other hand, spatial reuse, also known as frequency reuse, is a powerful technique to improve the area spectral efficiency of multi-user communication systems. Cellular systems are examples whereby radios exploit the power falloff with distance and reuse the same frequency for transmission at spatially separated locations\cite{spatialcell1}. Similar ideas can be applied to users in a distributed wireless network, where different transmit-receive (Tx-Rx) pairs at a distance away are allowed to transmit simultaneously, with the objective to achieve higher system capacity whilst still meeting all the transmission quality requirements\cite{spatialreuse}\cite{spatialpower}. In this paper, the Aloha game model in \cite{alohagames} is generalized to include the spatial reuse capability, named as a \textit{generalized Aloha game}. Unlike the model in \cite{alohagames}, the use of spatial reuse here distorts the symmetric structure in the expressions to evaluate the NE solution. As a result, a new Lyapunov function needs to be constructed to prove the convergence of a generalized Aloha game.

Also notice that our generalized Aloha game is different from the MPR Aloha game model in \cite{MacKenzieMPR}, although they both allow multiple packets to be successfully received simultaneously. In \cite{MacKenzieMPR}, the users are assumed to be indistinguishable; every user knows the current number of backlogged users in the system; the MPR model is possible by enabling multiple captures in a single channel, or single capture via the use of multiple parallel channels; the stability conditions and stability region of the equilibrium strategy are based on the drift analysis of a Markov chain and the selfish behavior of users when the number of backlogged users goes to infinity.  On the other hand, the generalized Aloha game in this paper assumes that users are heterogeneous in their bandwidth requirements and their neighboring user environment; users have information about the transmission probabilities of others; the MPR capability comes from the spatial reuse of a conventional Aloha collision channel; the stability issues are based on the Lyapunov stability analysis of a nonlinear system. These differences would be further discussed when we introduce the interference matrix in our Aloha game model in Section \ref{SectionModel}.

We introduce the model for the generalized Aloha game in Section \ref{SectionModel}, follow by some mathematical fundamentals on fixed point theory and
order theory in Section \ref{SectionMath}. We next discuss the existence of a Nash equilibrium (NE) in Section \ref{SectionNE}. In particular, we use fixed
point theory and order theory to prove the existence of a least fixed point in the generalized Aloha game, which is the unique NE of the game and the most
energy-efficient operating point for all players. In Section \ref{SectionStability}, we propose a method to prove the stability of the NE. The Krasovskii's
method is used to construct the Lyapunov function and obtain the conditions to examine the stability of the NE. After obtaining the conditions to test for
system stability, we summarize how to dynamically converge to the least fixed point in game iterations. Section \ref{SectionSimulation} shows through
simulations that the generalized Aloha game is applicable to large-scale distributed systems with complicated network topologies.  An empirical
relationship between the achievable total throughput and the network connectivity is finally obtained through our simulations. We conclude the paper in
Section \ref{SectionConclusion}.

\section{Model for Aloha Games with Spatial Reuse}\label{SectionModel}
Consider a distributed network with $N$ transmitters, where each transmitter has its unique designated receiver.
Each Tx-Rx pair is a player who competes for the channel to transmit. The conventional Aloha games are generalized to the scenarios where there exists spatial reuse among a group of non-cooperative players, i.e., those players who will not interfere each other can transmit concurrently. Here, only a connected network is considered (If the network is not connected, then it can be divided into several independent connected sub-networks, and then be dealt with separately). We assume that every player's transmission queue is continuously backlogged, i.e., the transmitter of every player always has a packet to transmit to its designated receiver.

\begin{figure}[!t]
\centering
\includegraphics[width=0.4\linewidth]{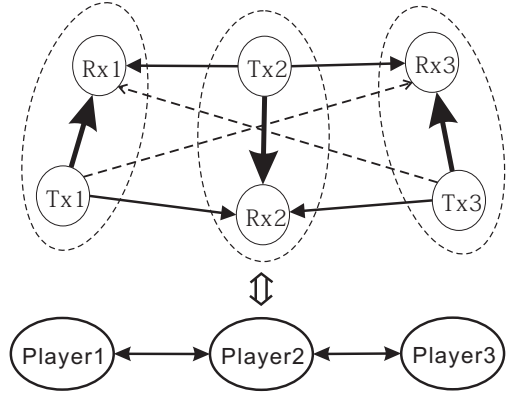} 
\caption{The chain-like topology for 3 transmit-receive pairs} \label{pair}
\end{figure}

As an example, three Tx-Rx pairs and their equivalent chain-like topology are shown in Fig. \ref{pair}, where players 1 and 3 can transmit concurrently
without collisions but neither of them can transmit together with player 2. Such interference relations can be characterized by an interference matrix
\textbf{A}. For the chain-like topology given in Fig.1,
\[
\textbf{A} = \begin{bmatrix}
0 & 1 & 0\\
1 & 0 & 1\\
0 & 1 & 0
\end{bmatrix}
 \]%
in which $a_{12} = 1$ means player 2 is a one-hop neighbor of player 1, $a_{13} = 0$ means player 3 is not a one-hop neighbor of player 1, etc. Notice that
in this example $\textbf{A}$ is a symmetric matrix. However, $ a_{ij}=a_{ji} $ is not necessarily true, i.e., the interference topology is a directed
graph.

The interference matrix characterizes the spatial distribution and frequency reuse capability of the players. Each player has different neighboring players
which directly affect its transmission. Such interference relations between the players cannot be properly described in the form of the MPR matrix in
\cite{MacKenzieMPR}. The multi-packet reception (MPR) model in \cite{MacKenzieMPR} is defined by a MPR matrix $\textbf{R}$, whose entries
$\rho_{nk}=[\textbf{R}]_{nk}$ is defined as the probability that $ k $ packets are successfully received in a slot when $ n $ packets are transmitted. The
MPR matrix is used to characterize the MPR model in a probabilistic manner, in which all users are assumed to be indistinguishable and have equal chance to
transmit successfully. In contrast, the spatial reuse considered in the interference matrix is deterministic and specific to each player, whose best
response would then be tailored to its specific neighboring environment.

With the defined interference matrix, we can now study the behavior of a generalized Aloha game. The objective of the game is for player $i$ to select a
suitable transmission probability $q_i$ so that player $i$ achieves its target rate $ y_i $, $\forall i \in \mathcal{N}= \{1, 2, \cdots, N\}$, with the
lowest possible energy consumption, i.e., each player uses the smallest transmission probability as it could to attain its target rate. The target rate
combination $\underline{y}=[y_1,\cdots,y_N]$ is controlled by certain pricing strategies \cite{alohaprice} or some commonly agreed adjusting rules that try
to achieve Pareto efficiency \cite{Pareto}. When the target rate $y_i$ is achieved, we have
\begin{equation}\label{ideal}
y_i = q_i \prod_{a_{ij}=1}(1-q_j), \qquad \forall i \in \mathcal{N}.
\end{equation} %

The equation indicates that for a successful transmission for player $i$, all those players which will interfere with its transmission (player $j$ where
$a_{ij}=1)$, should not transmit. It can be seen that the equations obtained here do not have a symmetric structure since the transmission probabilities of
some of the players are missing in some of the equations, depending on the network interference topology. This is unlike the relationship obtained from a
fully connected network\cite{alohagames}, where $a_{ij}=1, \forall i\ne j$; in this situation, the equations exhibit symmetric structures. We now formally
state the generalized Aloha game as follows:

\textit{Players}: Distributed Tx-Rx pairs, $i\in \mathcal{N}$, who compete for a single collision channel to transmit via slotted-Aloha-like random access
scheme.

\textit{Actions}: Each player \textit{i} chooses a transmission probability $q_i\in [0,1]$, $\forall i \in \mathcal{N}$.

\textit{Objectives}: Each player \textit{i} ($i \in \mathcal{N}$) aims to minimize the energy consumption in attaining its target rate $y_i$, i.e.,

\begin{equation}\label{OptimizationProblem}
\begin{array}{l l}
\min ~~ q_i\\
\mbox{s.t.} ~~ \ y_i = q_i \prod_{a_{ij}=1}(1-q_j).
\end{array}
\end{equation}

The solution of a generalized Aloha game is \textit{Nash equilibrium} (NE), which is defined as an action profile (in our case,
$\underline{q}^*=[q_1^*,\cdots,q_N^*]$) in which each action is a best response to the actions of all the other players\cite{Nash}. To be qualified as a NE
of the generalized Aloha game, the constraints of all players have to be satisfied first, or in other words, we can directly examine the solutions to the
set of equations in (\ref{ideal}). Interesting questions arise related to the problem. By definition, since all the transmission probabilities are
real-valued and cannot exceed 1, some of the solutions to (\ref{ideal}) that do not satisfy these constraints should be discarded. Among the remaining
solutions, is there an optimal one existing for all players, or multiple solutions each favoring different subgroups of players? In case if there are
multiple solutions to (\ref{ideal}), how to make the players reach the consensus to choose the same solution? If the consensus can be made, how to
dynamically reach that solution in game iterations?

Suppose the iterative approach in \cite{alohagames} is applied to (\ref{ideal}) to find out the solutions, i.e., the transmission probability of player \textit{i} at the ($ m+1 $)th iteration of the game is given by
\begin{equation}\label{iteration}
q_i^{(m+1)} = \min\{\frac{y_i}{\prod_{ a_{ij}=1}(1-q_j^{(m)})}, 1\}, \quad \forall i\in \mathcal{N}.
\end{equation}
If a solution $\underline{q}_s =[q_{s,1}, q_{s,2}, \cdots, q_{s,N}]$ exists, it should satisfy
\begin{equation}\label{fixedpointSpatial}
  q_{s,i} = \min\{\frac{y_i}{\prod_{a_{ij}=1}(1-q_{s,j})}, 1\}, \quad \forall i\in \mathcal{N}.
\end{equation}

Besides satisfying the equality constraints in (\ref{ideal}), if there exist multiple feasible solutions, we will prove in Section \ref{SectionNE} that
there exists an optimal solution which enable each player to operate with the minimal transmission probability. This optimal solution $\underline{q}^*$ is
then the unique NE of the generalized Aloha game defined in (\ref{OptimizationProblem}). Mathematically, if we introduce a binary relation ``$ \preceq $''
between two real-valued vectors $ \underline{a},\underline{b}$, which is defined as component-wise less than or equal to, i.e., $ \underline{a}\preceq
\underline{b}\Leftrightarrow a_i\leq b_i,\forall i\in \mathcal{N} $, then the NE $\underline{q}^*$, as compared to other solutions $\underline{q}_s$, would
satisfy $ \underline{q}^*\preceq \underline{q}_s$.

\section{Mathematical Foundation}\label{SectionMath}
In this section, we introduce the Brouwer's fixed-point theorem (in order to prove the existence of solutions to (\ref{fixedpointSpatial})), the Kleene
fixed-point theorem (in order to prove the existence of a least fixed point, which is later shown to be the unique NE of the game), and some related
definitions. These proofs involve an \textit{N}-dimensional vector function $ \underline{F}=(f_1, f_2, \cdots, f_N)^T$, whose component
$f_i(\underline{q}) (i\in \mathcal{N})$ is a real-valued function of $ \underline{q}\in K$, where $K\subset \mathbb{R}^N $.

\newtheorem{thm}{Theorem}
\begin{thm}[Brouwer's fixed-point theorem \cite{brouwer}]
Every continuous vector function \underline{F} from a convex compact set $K$ (where $K \subset \mathbb{R}^N$) to $K$ itself has a fixed point, i.e., there is a point $ \underline{q} \in K$ such that $ \underline{F}(\underline{q})=\underline{q} $.
\end{thm}

\newtheorem{dfn}{Definition}
\begin{dfn}\label{binary}
A \textbf{binary relation} over a set $K\subset \mathbb{R}^N$ is a collection of ordered pairs in $K$.
\end{dfn}

\begin{dfn}\label{partial}
A binary relation ``$ \preceq $'' over a set $K\subset \mathbb{R}^N$ is a \textbf{partial order} if it is reflexive, antisymmetric, and transitive, i.e., $ \forall \underline{a},\underline{b},\underline{c}\in K $,

(a) reflexivity: $\underline{a}\preceq \underline{a}$;

(b) antisymmetry: if $ \underline{a}\preceq \underline{b} $ and $ \underline{b}\preceq \underline{a} $, then $ \underline{a}=\underline{b} $;

(c) transitivity: if $ \underline{a}\preceq \underline{b} $ and $ \underline{b}\preceq \underline{c} $, then $ \underline{a}\preceq \underline{c} $.
\end{dfn}

\begin{dfn}\label{directed}
A subset \textit{S} of a partially ordered set ($K$, $ \preceq $) is called \textbf{directed} if, for any $ \underline{a},\underline{b}\in S$, there is $ \underline{c} \in S $ such that $
\underline{a}\preceq \underline{c} $ and $ \underline{b} \preceq \underline{c} $.
\end{dfn}

\begin{dfn}\label{complete}
A partially ordered set ($K$, $ \preceq $) is said to be \textbf{complete}, and hence a \textbf{complete partial order}, if there is a least element of $K$
(denoted by $\perp$) and every directed subset $ S\subset K $ has a least upper bound $ \sup S \in K$.
\end{dfn}

\begin{dfn}\label{monotonic}
Let ($K$, $ \preceq $) be a partially ordered set. A vector function $ \underline{F}:K\to K $ is \textbf{monotonic} or \textbf{order-preserving} if whenever $ \underline{a}\preceq
\underline{b} $, we have $ \underline{F}(\underline{a})\preceq \underline{F}(\underline{b}) $.
\end{dfn}

\begin{dfn}\label{Scott}
Given a partially ordered set ($K$, $ \preceq $), a vector function $ \underline{F}:K\to K $ is \textbf{Scott-continuous} if, for every directed subset \textit{S} of $K$, $ \sup
\underline{F}(S)=\underline{F}(\sup S)\in K $.
\end{dfn}

\begin{thm}[Kleene fixed-point theorem \cite{Kleene}\cite{FixedPointTheorems}]
Let ($K$, $ \preceq $) be a complete partial order, and let $ \underline{F}:K\to K $ be a Scott-continuous vector function. Then \underline{F} has a least
fixed point, which is the supremum of the ascending Kleene chain of \underline{F}.
\end{thm}

The ascending Kleene chain of $\underline{F}$ is the chain
\[
\perp \preceq \underline{F}(\perp) \preceq \underline{F}(\underline{F}(\perp))\preceq \cdots \preceq \underline{F}^n(\perp)\preceq \cdots
 \]%
obtained by iterating $\underline{F}$ on the least element $ \perp $ of $K$.

Expressed in a formula, the theorem states that

\begin{equation}
\textbf{LFP}(\underline{F})=\sup_{n\rightarrow \infty}\underline{F}^n(\perp)
\end{equation}
where \textbf{LFP} denotes the least fixed point, which is less than or equal to all other fixed points by some partial order.

\section{Equilibrium of the generalized Aloha Game}\label{SectionNE}
In this section, we would prove the existence of solutions to (\ref{fixedpointSpatial}), and the existence of a least fixed point which would later be shown to be the unique NE in the generalized Aloha game. We specify the aforementioned \textit{N}-dimensional vector function $ \underline{F}=(f_1, f_2, \cdots, f_N)^T$, whose component $f_i$ is defined as a real function given by
\begin{equation}\label{fn}
f_i(\underline{q}) = \min\{\frac{y_i}{\prod_{a_{ij}=1}(1-q_j)}, 1\}, \quad \forall i\in \mathcal{N}.
\end{equation}
The function $f_i$ maps $\underline{q}=[q_1, \cdots, q_N] \in [0,1]^N$ into the $i$th component of the vector function $\underline{F}$. The reason to define (\ref{fn}) will become clearer shortly.

\subsection{Existence of Solutions}
The equations defined in (\ref{fn}) and Brouwer's fixed-point theorem are used to examine the existence of solutions to (\ref{fixedpointSpatial}). From
definition, the fixed point of the vector function $\underline{F}$, $\underline{q}_s =(q_{s,1},q_{s,2},\cdots,q_{s,N})$, is given by solving
$\underline{F}(\underline{q}_s)=\underline{q}_s$, or $ f_i(\underline{q}_s)=q_{s,i},\forall i \in \mathcal{N} $. By substituting such a relationship into
(\ref{fn}), it would result in the solution having the same form as that obtained in (\ref{fixedpointSpatial}). This means that the solution to
(\ref{fixedpointSpatial}) can be understood as a fixed point to the defined vector function $ \underline{F}$. Since the continuous vector function $
\underline{F} $ maps a point $ \underline{q} $ from the convex compact set $ K\equiv [0,1]^N $ to $K$ itself, according to Brouwer's fixed-point theorem,
there exists a point $ \underline{q}_s $ such that $ \underline{q}_s=\underline{F}(\underline{q}_s) $, i.e., (\ref{fixedpointSpatial}) follows.

It is now clear why (\ref{fn}) is defined, as the fixed point behavior of (\ref{fn}) is equivalent to the original equality constraints defined in (\ref{ideal}) except that we explicitly include the bound $q_i=1$ in (\ref{fn}) to ensure that $\underline{F}$ maps into a compact set. One issue to take note is that by using (\ref{fn}) to replace (\ref{ideal}), an extraneous solution $\underline{q}_s=\underline{1} $ has been introduced to the original equality constraints defined in (\ref{ideal}). This solution is not desirable since all players continuously transmit and all transmissions will result in contention. Fortunately, this undesirable solution to (\ref{ideal}) can be easily identified and discarded.

Since the fixed points are proper only if they exist in $(0,1)^N$, we focus on such solutions in the following discussion. We will show that if multiple
fixed points exist in $(0,1)^N$, there should exist a most energy-efficient one.

\subsection{Existence of a Least Fixed Point}

In the discussion for the generalized Aloha game, the following properties about ``$ \preceq $'' over the set $K$ hold:

(a) By Definitions \ref{binary} \& \ref{partial}, the binary relation ``$ \preceq $'' over the set $K=[0,1]^N$ is a partial order, since it is reflexive, antisymmetric, and transitive.

(b) By Definitions \ref{directed} \& \ref{complete}, the partially ordered set ($K$, $\preceq$) is a complete partial order. The least element of $K$ is given by \underline{0}. For every directed subset $ S\subset K $, the least upper bound of \textit{S} is the largest element in \textit{S}, thus $ \sup S \in K $. Therefore, ($K$, $\preceq$) is a complete partial order.

We are now ready to prove the following propositions.

\newtheorem{proposition}{Proposition}
\begin{proposition}
The vector function $ \underline{F} $ in (\ref{fn}) is an order-preserving function over the complete partial order ($K$, $\preceq$).
\end{proposition}

For any two vectors $ \underline{q}, \underline{p}\in K $, where $ \underline{q}\preceq \underline{p} $, we have
\[
\min\{\frac{y_i}{\prod_{a_{ij}=1}(1-q_j)}, 1\} \leq \min\{\frac{y_i}{\prod_{a_{ij}=1}(1-p_j)}, 1\} ,
\]
for $i \in \mathcal{N}$. According to Definition \ref{monotonic}, the proposition holds.

\begin{proposition}
For the vector function $ \underline{F} $ defined by (\ref{fn}), if there exist multiple fixed points in $(0,1)^N$, then a least fixed point exists, which is less than or equal to
all other fixed points, according to the partial order ``$ \preceq $'' over the set $K$.
\end{proposition}

According to Definition \ref{Scott}, the vector function $ \underline{F} $ defined by (\ref{fn}) is Scott-continuous, because for every directed subset \textit{S} of $K$, $ \sup \underline{F}(S)=\underline{F}(\sup S)\in K $, which follows from the order-preserving properties of $ \underline{F} $.

In summary, by Kleene fixed-point theorem, the vector function $ \underline{F} $ defined by (\ref{fn}) has a least fixed point, which is less than or equal to all other fixed points, in the partial order ``$ \preceq $''. Moreover, the least fixed point can be obtained by iterating $ \underline{F} $ on the least element of $K$, i.e.,
\begin{equation}\label{lfp1}
\textbf{LFP}(\underline{F})=\sup_{n\rightarrow \infty}\underline{F}^n(\underline{0}).
\end{equation}%

\subsection{Initialization}\label{Initialization}

Eq.(\ref{lfp1}) suggests that the players can choose initial transmission probabilities $ \underline{q}^{(0)}=\underline{0} $ to reach the least fixed point by game iteration.
Actually we are able to prove that the initial transmission probabilities $ \underline{q}^{(0)}$ can be set as any point in the set $I\equiv[0,y_1]\times[0,y_2]\times\cdots[0,y_N]$, i.e., $\underline{q}^{(0)} \preceq \underline{y}$. This can be done using Kleene fixed-point theorem. By replacing $K\equiv [0,1]^N$ with $K'\equiv[q_1^{(0)},1]\times[q_2^{(0)},1]\times\cdots[q_N^{(0)},1]$, one can easily use the earlier approach to show that $K'$ is convex and compact, and can verify that ($K'$, $\preceq$) is a complete partial order. Moreover, from the structure of the continuous vector function $ \underline{F} $ defined by (\ref{fn}), one easily sees that $\underline{F}(\underline{q})\succeq \underline{y} \succeq \underline{q}^{(0)}$, therefore $\underline{F}$ maps a point $\underline{q}$ from $K'$ to $K'$ itself. Finally, one can also verify that $\underline{F}: K'\rightarrow K'$ is a Scott-continuous vector function. Therefore, by Kleene fixed-point theorem, the least fixed point can be obtained by iterating $\underline{F}$ on the least element of $K'$, i.e.,
\begin{equation}\label{lfp}
\textbf{LFP}(\underline{F})=\sup_{n\rightarrow \infty}\underline{F}^n(\underline{q}^{(0)}), \underline{q}^{(0)} \in I.
\end{equation}

Notice that the actual feasible region for initial transmission probabilities is larger than $I$. Later in Section VI.A we will numerically show that, for a stable NE, there exists a neighborhood $\Omega$ of this NE such that the system starting from any point in $\Omega$ will converge to this NE.
However, while the value of the least fixed point is not known at the point of evaluating, it might be sufficient to look for initial probabilities just from the region $I$.

\subsection{Discussion}

The existence of a least fixed point is of great significance. If there exist multiple fixed points (i.e., multiple solutions to (\ref{ideal})) in $K$, every selfish player will choose the fixed point which is best for itself. If the least fixed point exists, then the transmission probability for every player will be the least at this point, thus this fixed point is also the most energy-efficient for every player. As a result, every player will choose this fixed point as the operating point. Therefore, the least fixed point is the unique NE of the generalized Aloha game.
Finally, we have proved that the players can choose any initial transmission probabilities $ \underline{q}^{(0)}\in I $ to reach the least fixed point by game iteration. These results have not been pointed out in the existing work published in \cite{alohagames}\cite{alohaprice}, where the Aloha game model is first brought up.

On the other hand, (\ref{lfp}) does not guarantee that such an iteration process is stable, i.e., the solution may still diverge due to small disturbance at this fixed point. In the next section, we will discuss the method to prove the stability of the NE.

\section{Stability of the Equilibrium Point}\label{SectionStability}
This section investigates the stability of a generalized Aloha game defined by the iteration process in (\ref{iteration}). Stability is a desired property
of the NE. A stable NE can absorb small disturbances within a certain neighborhood $\Omega$, e.g., due to the inaccuracy in estimating other players'
transmission probabilities. On the other hand, if a NE is not stable, then the game iteration process will diverge to some undesirable states such as
$\underline{q}=\underline{1}$, which unfortunately leads to network congestion and results in zero throughputs for everyone. Another motivation is that by
understanding the conditions to maintain network stability, we hope to acquire good knowledge to design the intelligence inside future self-autonomous
radios. We will give a discussion on this in section \ref{SectionSimulation} using an example.

To prove the stability of the resulting NE, we follow the pattern from \cite{alohagames} and approximate the generalized Aloha game by the Jacobi update scheme:
\begin{equation}\label{update}
\underline{q}^{(m+1)} = \underline{q}^{(m)} + \epsilon (\underline{F}(\underline{q}^{(m)})-\underline{q}^{(m)})
\end{equation}%
where $\epsilon$ is a fixed small positive number and $\underline{F}$ is defined by (\ref{fn}). For sufficiently small $\epsilon$, (\ref{update}) can be
approximated by a continuous-time game:
\begin{equation}\label{sys1}
\underline{\dot{q}}(t) = \underline{g}(\underline{q}(t)) = \underline{F}(\underline{q}(t))-\underline{q}(t)
\end{equation}%

In the presence of spatial reuse, functions defined in (\ref{fn}) do not have a symmetric structure since the transmission probabilities of some of the
players are missing in some of the equations, depending on the network interference topology. As a result, the Lyapunov function $ \Lambda(\underline{q}) $
in \cite{alohagames} is no longer applicable to the scenarios with spatial reuse. Therefore, it is necessary to develop a more general Lyapunov function to
examine the stability of the solutions.

\subsection{Krasovskii's Method}
We use a new method to construct a Lyapunov function to prove system stability, namely the Krasovskii's method\cite{gradient}.
\begin{thm}[Krasovskii's Method]
Consider the non-linear system defined by $\dot{\underline{x}}=\underline{g}(\underline{x})$, with the equilibrium point of interest being the origin. Let $\textbf{J}(\underline{x})$ denote the Jacobian matrix of the system, i.e., $\textbf{J}(\underline{x})=\partial \underline{g} / \partial \underline{x}$. If the matrix $\textbf{B}(\underline{x})=\textbf{J}(\underline{x})+\textbf{J}^T(\underline{x})$ is negative definite in a neighborhood $\Omega$, then the equilibrium at the origin is asymptotically stable. A Lyapunov function for this system is given by $\Lambda(\underline{x})=\underline{g}^T(\underline{x})\underline{g}(\underline{x})$.
\end{thm}

Define $\textbf{C}(\underline{q})=-\textbf{B}(\underline{q})=-[\textbf{J}(\underline{q})+\textbf{J}^T(\underline{q})]$, where $ \textbf{J}(\underline{q}) $ is the Jacobian matrix of the system in (\ref{sys1}). For those fixed points in $ (0,1)^N $, the entries of $ \textbf{J}(\underline{q}) $ can be calculated as follows:
\begin{equation}\label{entries}
\begin{array}{l l}
[\textbf{J}(\underline{q})]_{ij}=[\partial \underline{g} / \partial \underline{q}]_{ij}=\frac{\partial g_i}{\partial q_j}=\\
\left\{
  \begin{array}{l l}
    -1 & \quad \textrm{ $i=j$}\\
    0 & \quad \textrm{ $i\neq j, a_{ij}=0$}\\
    \frac{f_i(\underline{q})}{1-q_j} & \quad \textrm{ $i\neq j, a_{ij}=1$}\\
  \end{array} \right.
  =\left\{
    \begin{array}{l l}
      -1 & \quad \textrm{ $i=j$}\\
      \frac{a_{ij}f_i(\underline{q})}{1-q_j} & \quad \textrm{ $i\neq j$}\\
    \end{array} \right.
\end{array}
\end{equation}

Notice that at a fixed point in $ (0,1)^N $, $ \underline{\dot{q}}(t) =\underline{g}(\underline{q}(t))= \underline{F}(\underline{q}(t))-\underline{q}(t)=0 $, i.e.,
\begin{equation}\label{Equilibrium}
q_{s,i}=f_i(\underline{q}_s) = \frac{y_i}{\prod_{a_{ij}=1}(1-q_{s,j})}.
\end{equation}

Therefore, the entries of $ \textbf{C}(\underline{q}_s) $ at the fixed point can be obtained from (\ref{entries}) and (\ref{Equilibrium}):
\begin{equation}\label{C}
\begin{array}{l l}
[\textbf{C}(\underline{q}_s)]_{ij}=-[\textbf{J}(\underline{q}_s)+\textbf{J}^T(\underline{q}_s)]_{ij}\\
=\left\{
    \begin{array}{l l}
      2 & \quad \textrm{ $i=j$}\\
      -\frac{a_{ij}q_{s,i}}{1-q_{s,j}}-\frac{a_{ji}q_{s,j}}{1-q_{s,i}} & \quad \textrm{ $i\neq j$}\\
    \end{array} \right.
\end{array}
\end{equation}

An equivalent condition for the positive definiteness of $\textbf{C}(\underline{q}_s)$ is stated in Lemma 1\cite{matrix}.
\theoremstyle{plain}
\newtheorem{lem}{Lemma}
\begin{lem}
The real-valued square matrix $\textbf{C}_{N\times N}  $ is positive definite if and only if $ \det \textbf{C}_i>0 $ for $ i=1,2,\cdots,N $, where $ \textbf{C}_i $ is the leading principal sub-matrix of \textbf{C} determined by the first i rows and columns.
\end{lem}

Alternatively, a sufficient condition for the positive definiteness of $\textbf{C}(\underline{q}_s)$ is that $\textbf{C}(\underline{q}_s)$ be diagonally dominant\cite{matrix}, i.e.,
\begin{equation}\label{sufficient}
\sum\limits_{j=1}^{N}(\frac{a_{ij}q_{s,i}}{1-q_{s,j}}+\frac{a_{ji}q_{s,j}}{1-q_{s,i}})<2, \quad \forall i\in \mathcal{N}.
\end{equation}

If $\textbf{C}(\underline{q}_s)$ is positive definite, according to the Krasovskii's method, the following proposition holds with the corresponding Lyapunov function being $ \Lambda(\underline{q})= \underline{g}^T(\underline{q})\underline{g}(\underline{q})$.

\begin{proposition}\label{stable}
If there is a fixed point $ \underline{q}_s\in [0,1]^N$ with $\textbf{C}(\underline{q}_s)$ being positive definite, then there is a neighborhood
$\Omega \subset [0,1]^N$ of $\underline{q}_s$ such that: for any initial transmission probabilities $\underline{q}(0)\in \Omega$, the function
$\underline{q}(t)$ obeying the dynamics (\ref{sys1}) will converge to $\underline{q}_s\in \Omega$ as $t\to \infty$.
\end{proposition}

For sufficiently small $\epsilon$, Proposition \ref{stable} can be adjusted to make a statement about the convergence of (\ref{update}) at a fixed point
$\underline{q}_s$. Following the postulation in \cite{alohagames}, we also postulate that the fixed points that are stable for (\ref{sys1}) are also stable
when $\epsilon=1$, i.e., for the original iteration (\ref{iteration}).

In summary, we can verify the positive definiteness of $
\textbf{C}(\underline{q}^*) $ in order to judge the stability of the NE $\underline{q}^*$ in a generalized Aloha game. Certain necessary and sufficient conditions could be used, e.g., Lemma 1. Alternatively, the sufficient condition given in (\ref{sufficient}) can also be used, which is easier to implement and gives almost the same bound.

\subsection{Stability Comparison between Multiple Fixed Points}
Stability is a desired property of the NE. The price of instability is that the whole network would be congested and nobody can transmit successfully.
This subsection would compare the stability of the fixed points. Specifically, the following proposition suggests that the least fixed point is more likely to be stable than other fixed points. Therefore, the least fixed point not only is optimal in terms of energy efficiency, but also carries less risk of instability.
\begin{proposition}
If the least fixed point is not stable in Aloha games, nor are other fixed points.
\end{proposition}

\textit{Proof:} Let's investigate the entries of $\textbf{C}(\underline{q}_s)$ from (\ref{C}) first.
\begin{equation}
c_{ij}=[\textbf{C}(\underline{q}_s)]_{ij}=\left\{
    \begin{array}{l l}
      2 & \quad \textrm{ $i=j$}\\
      -\frac{a_{ij}q_{s,i}}{1-q_{s,j}}-\frac{a_{ji}q_{s,j}}{1-q_{s,i}} & \quad \textrm{ $i\neq j$}\\
    \end{array} \right.
\end{equation}

When $i\neq j$, $ c_{ij}$ is a non-increasing function of $ \underline{q}_s $ in the partial order ``$ \preceq $'', i.e., if two fixed points satisfy $ \underline{q}_s\preceq  \underline{p}_s $, then $ c_{ij}(\underline{q}_s)\ge c_{ij}(\underline{p}_s), \forall i\neq j$.

Define a function $ h(\underline{x},\underline{q}_s) $ as follows:
\begin{equation}
h(\underline{x},\underline{q}_s)=\underline{x}^T\textbf{C}(\underline{q}_s)\underline{x}=\sum 2x_i^2+\sum\limits_{i\neq j}^{}c_{ij}(\underline{q}_s)x_ix_j,
\end{equation} %
where $ \underline{x}\in \mathbb{R}^N $ is the variable, and $ \underline{q}_s $ is a parameter.

Now suppose the least fixed point $ \underline{q}^* $ is not stable, i.e., $ \textbf{C}(\underline{q}^*) $ is not positive definite. According to the definition of positive definiteness\cite{matrix}, $ \exists \underline{x}\neq \underline{0},\underline{x}\in \mathbb{R}^N$, such that $h(\underline{x},\underline{q}^*)\leq 0 $.

Since $ c_{ij}\leq 0, \forall i\neq j $, we can always find an $ \underline{x} $ with $ x_ix_j\geq 0, \forall i\neq j $, such that $h(\underline{x},\underline{q}^*)\leq 0 $. For such a given $ \underline{x} $ and any other fixed point $ \underline{p}_s\succeq \underline{q}^* $, we have
\begin{equation}
h(\underline{x},\underline{q}^*)-h(\underline{x},\underline{p}_s)=\sum\limits_{i\neq j}^{}x_ix_j(c_{ij}(\underline{q}^*)-c_{ij}(\underline{p}_s))\ge 0.
\end{equation}

Consequently, $ h(\underline{x},\underline{p}_s)\leq h(\underline{x},\underline{q}^*)\leq 0 $, i.e., $ \textbf{C}(\underline{p}_s) $ is not positive definite. Therefore, if the least fixed point is not stable, nor are other fixed points.$\blacksquare$

In summary, we only need to focus on the stability of the least fixed point. If it is not stable, then no stable equilibrium point exists; if it is stable, then it will be the choice of all players, and the behavior of the remaining fixed points may not be of our concern since they are not energy efficient even if the solution is stable.
The reason behind this can be easily interpreted as follows. If each player transmits more often but achieves the same throughput, it is just an indication that it is likely there are more collisions in the network and hence more likely that the network will become congested.

\subsection{How to Dynamically Converge to the Least Fixed Point}
We summarize our results. First, we construct an interference matrix \textbf{A} based on a given distribution of players. Second, for a given target rate
combination $ \underline{y}=[y_1,\cdots,y_N] $, we iteratively calculate the least fixed point $ \underline{q}^* $ of the vector function $ \underline{F} $
defined in (\ref{fn}) by choosing the initial point $ \underline{q}^{(0)}\in I $. Third, we judge the stability of $ \underline{q}^* $ by verifying the
positive definiteness of the matrix $ \textbf{C}(\underline{q}^*) $ given in (\ref{C}), based on Proposition \ref{stable}. Finally, if the least fixed
point $ \underline{q}^* $ is stable, then all players can arrive at this equilibrium point through iterations, by choosing any initial point from the set
$I$.

In short, for a combination of target rates satisfying the stability conditions given by Proposition \ref{stable}, all players can reach the least fixed
point (i.e., the unique NE of the game) as a stable operating point, by choosing any initial transmission probabilities $ \underline{q}^{(0)}\in I $, among
which $\underline{0}$ and $\underline{y}$ are two convenient choices.

\section{Simulation Studies}\label{SectionSimulation}

In Part A of this section, we first demonstrate the existence of the least fixed point and the use of the Krasovskii's method to check its stability by
using the three-player chain-like topology. The actual iteration process is simulated so as to test the stability of the fixed points predicted by the
Krasovskii's method. The Region of Attraction (RoA) of the least fixed point is estimated by using the Lyapunov function. Then we study the behavior of the
fixed points with one varying parameter $y_i$. Moreover, the combinations of maximum achievable target rates for the players are plotted. Finally, we go
beyond the defined game and give simple illustrations on how future autonomous players can make use of the developed theory to improve the overall system
sum-rate. In Part B, our theory is applied to more complicated network topologies to examine the maximum achievable target rates, with the objective to
understand the relationship between the spatial reuse capability and the network connectivity.

\subsection{Three-player Chain-like Topology}
\subsubsection{Least Fixed Point}
We use the three-player chain-like topology in Fig. \ref{pair} as an illustration. Assume $ y_1=y_2=y_3=0.15 $, then the fixed points can be obtained by solving (\ref{ideal}),
which yields 3 solutions: $[q_{s,1},q_{s,2},q_{s,3}]$ = [0.1952, 0.2316, 0.1952], [0.5451, 0.7248, 0.5451], [1.4097, 0.8936, 1.4097]. Obviously the third solution is not feasible because two of the transmission probabilities are greater than 1. The first two solutions are in $[0,1]^3$ and are the feasible solutions to (\ref{ideal}). Denote the first solution as $\underline{q}^*$ and second solution as $\underline{p}_s$. It is obvious that $\underline{q}^*$ $\preceq$ $\underline{p}_s$. Therefore, $\underline{q}^*$ is the least fixed point.

\subsubsection{Krasovskii's Method}
The system dynamics is given by:
\begin{equation}
  \left\{
  \begin{array}{l l}
    \dot{q_1} = g_1(\underline{q}) = \min \{y_1/(1-q_2),1\}-q_1\\
    \dot{q_2} = g_2(\underline{q}) = \min \{y_2/(1-q_1)(1-q_3),1\}-q_2\\
    \dot{q_3} = g_3(\underline{q}) = \min \{y_3/(1-q_2),1\}-q_3\\
  \end{array} \right.
\end{equation}

The entries of $ \textbf{C}(\underline{q}_s) $ evaluated at a fixed point $\underline{q}_s$ can be obtained from (\ref{entries}) and (\ref{Equilibrium}):
\begin{equation}\label{positivedefinite}
\begin{array}{l l}
\textbf{C}(\underline{q}_s)=-[\textbf{J}(\underline{q}_s)+\textbf{J}^T(\underline{q}_s)]=\\
\begin{bmatrix}
2 & -\frac{q_{s,1}}{1-q_{s,2}}-\frac{q_{s,2}}{1-q_{s,1}} & 0\\
-\frac{q_{s,1}}{1-q_{s,2}}-\frac{q_{s,2}}{1-q_{s,1}} & 2 & -\frac{q_{s,2}}{1-q_{s,3}}-\frac{q_{s,3}}{1-q_{s,2}}\\
0 & -\frac{q_{s,2}}{1-q_{s,3}}-\frac{q_{s,3}}{1-q_{s,2}} & 2
\end{bmatrix}
\end{array}
\end{equation}

We now verify the stability of $\underline{q}^*$ and $\underline{p}_s$ using the Krasovskii's method by examining whether $ \textbf{C}(\underline{q}_s) $ given by (\ref{positivedefinite}) is positive definite. It can be claimed that $\underline{q}^*$ is stable while $\underline{p}_s$ is not.

\subsubsection{Game Iteration Process}
The iteration process of the generalized Aloha Game is given by (\ref{iteration}).
To verify the above claim about the stability of $\underline{q}^*$ and $\underline{p}_s$ using the Krasovskii's method, set $ y_1=y_2=y_3=0.15 $, set the initial transmission probabilities $ [q_1^{(0)},q_2^{(0)},q_3^{(0)}]$ equal to $[y_1,y_2,y_3] $ (P0), $\underline{q}^*$ and $\underline{p}_s$ separately, and run the process to see its actual performance.
\begin{figure}[!t]
\centering
\includegraphics[width=0.9\linewidth]{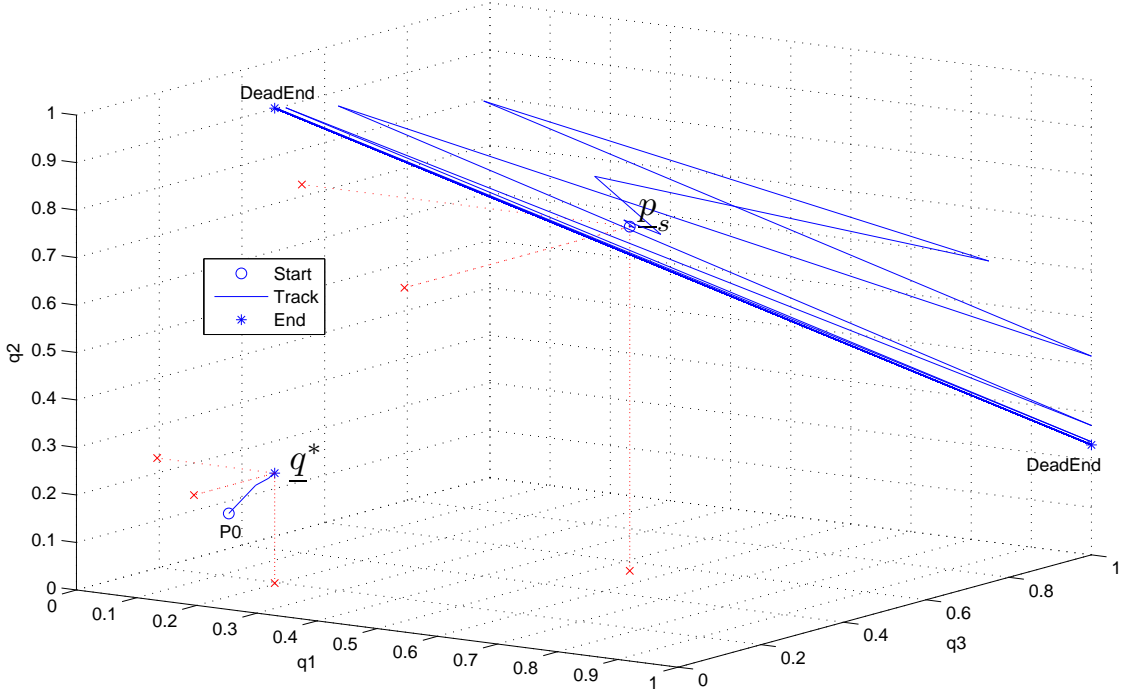}
\caption{The iteration process. P0 converge to $\underline{q}^*$, $\underline{p}_s$ is unstable} \label{track}
\end{figure}

We see from Fig. \ref{track} that the iteration starting at P0 converges to $\underline{q}^*$ within 10 iterations. On the other hand, we also see that the iteration starting at $\underline{p}_s$ ends up oscillating between two points, [0.1952, 1, 0.1952] and [1, 0.2316, 1]. Therefore, $\underline{q}^*$ is stable while $\underline{p}_s$ is not. This is consistent with the previous claim using the Krasovskii's method.

\subsubsection{Region of Attraction of the Least Fixed Point}
\textit{Region of Attraction (RoA)} is defined as the set of all initial points from which the system will converge to the equilibrium point as time goes to infinity\cite{RoA}. As was commented in \cite{RoA}, finding the exact RoA analytically might be difficult or even impossible. However, the Lyapunov function can be used to estimate the RoA. From Theorem 3 and Proposition 3, if the least fixed point is verified to be stable, then the neighborhood $\Omega$ specified using the Krasovskii's method is within the RoA. Therefore, for the three-player chain-like topology with $ y_1=y_2=y_3=0.15 $, we estimate the RoA for the least fixed point ( i.e., the NE $\underline{q}^*$), by verifying the positive definiteness of $\textbf{C}(\underline{q})$.
\begin{figure}[!t]
\centering
\includegraphics[width=0.9\linewidth]{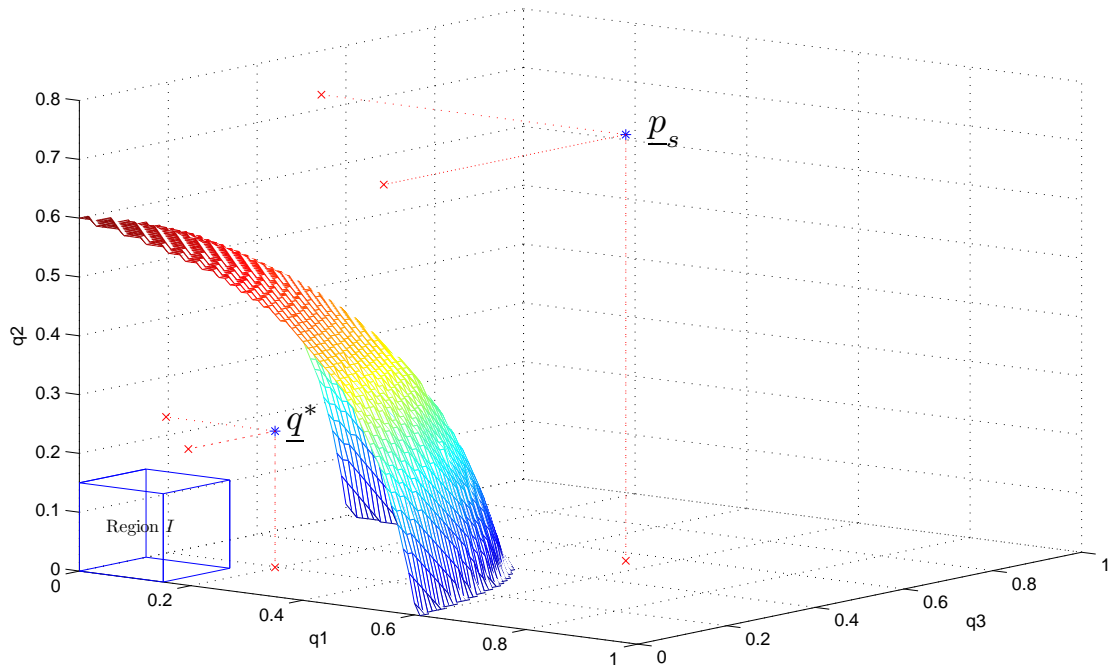}
\caption{Estimation of the region of attraction for the NE $\underline{q}^*$} \label{RoA}
\end{figure}

In Fig. \ref{RoA}, the region under the mesh surface provides an estimate of the RoA of the NE $\underline{q}^*$. Clearly the region $I$ (the cuboid near the origin) defined in Section \ref{Initialization} is within the RoA and can be obtained much easier. However, such an estimation is still quite conservative. For the described game iteration process, we actually observe that the set of points satisfying $\underline{q}\prec$ $\underline{p}_s$ are all within the RoA of the NE $\underline{q}^*$.

\subsubsection{Bifurcation of the Fixed Points}
\begin{figure}[!t]
\centering
\includegraphics[width=0.9\linewidth]{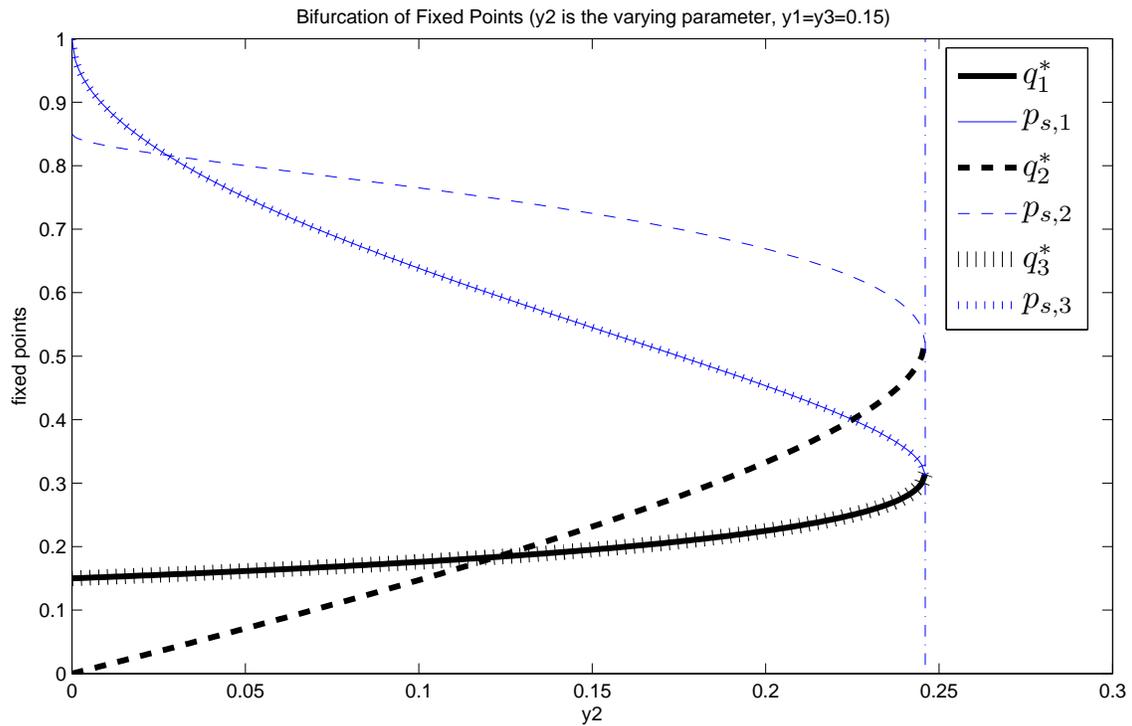}
\caption{Bifurcation of the Fixed Points ($y_2$ is the varying parameter, $y_1=y_3=0.15$)} \label{bifurcation}
\end{figure}

For the three-player chain-like topology, we study here the behavior of the fixed points with $y_2$ varying, while keeping $y_1=y_3=0.15$. For different values of $y_2$, we solve (\ref{ideal}), and plot the solutions accordingly in Fig. \ref{bifurcation} (denote the least fixed point as $\underline{q}^*$, the second fixed point as $\underline{p}_s$; the third solution is outside $[0,1]^3$).

From Fig. \ref{bifurcation} we observe that, as $y_2$ increases from 0 to 0.246, there exist two real-valued fixed points $\underline{q}^*, \underline{p}_s$ with $\underline{q}^*\preceq \underline{p}_s$. We verify using the Krasovskii's method that $\underline{q}^*$ is stable while $\underline{p}_s$ is not. When $y_2=0.246$, $\underline{q}^*$ and $\underline{p}_s$ coincide and obtain a critical equilibrium $\underline{q}_c^*=[0.3138,0.5223,0.3138]$, which corresponds to a zero eigenvalue of the Jacobian matrix $\textbf{J}(\underline{q}_c^*)$. If $y_2$ further increases, the fixed points disappear, i.e., there is no real-valued fixed point in $[0,1]^3$ (except the extraneous one $\underline{q}_s=\underline{1} $ introduced by including the bound $q_i=1$ in (\ref{fn})).
This phenomenon is mathematically named as \textit{Fold Bifurcation}\cite{fold}. This bifurcation is characterized by a single bifurcation condition that the Jacobian matrix $\textbf{J}(\underline{q}_c^*)$ has a codimension-one zero eigenvalue at the critical equilibrium point\cite{fold}.

Similar simulations with $y_1$ or $y_3$ being the varying parameter have been carried out, and we observe similar fold bifurcation of the fixed points.
Therefore, for the three-player chain-like topology, we postulate that at most one stable fixed point exists and it is the least fixed point.

We also extend the simulations to cases with more players and different topologies. Due to computational complexity of calculating all the solutions of (\ref{ideal}), we only examine cases with no more than 8 players. We observe three interesting phenomena: (1) there are at most two real-valued fixed points in $(0,1)^N$; (2) these two fixed points exhibit fold bifurcation with any of the target rate $y_i$ being selected as the varying parameter and the remaining fixed; (3) among these two fixed points, the least fixed point is stable while the other is not, before they coincide and disappear. However, a rigorous mathematical proof of such bifurcation behavior of the fixed points is still difficult, and might require further investigation.

In the next two subsections, we demonstrate how to extend the results obtained from the stability study and apply beyond the described game.

\subsubsection{Feasible Region of Target Rates}
We compare the maximum achievable target rates between a three-player chain-like topology and a fully connected topology (conventional Aloha games). We vary the combinations of the players' target rates $[y_1,y_2,y_3]$ and use the iterative approach to evaluate the least fixed point until the stability of this point cannot be achieved. We then plot the contour of $y_2$ for some given $[y_1, y_3]$.

\begin{figure}[!t]
\centering
\includegraphics[width=0.9\linewidth]{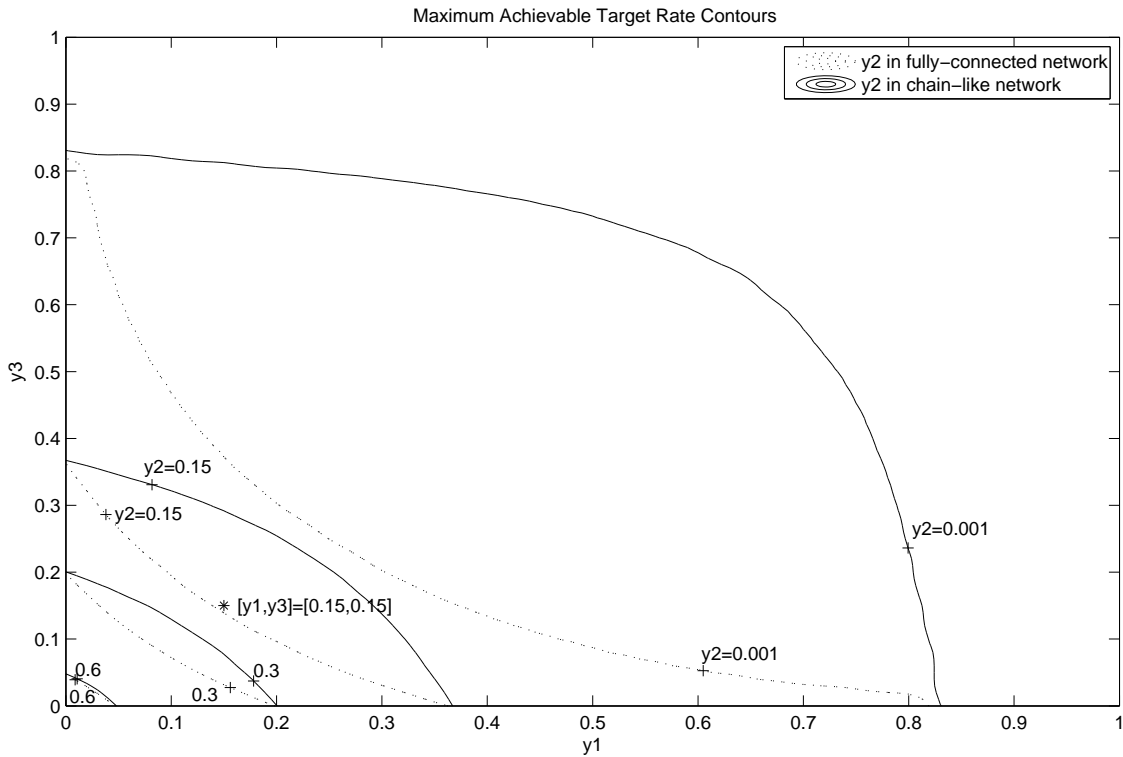}
\caption{Contour plot of the maximum achievable target rate} \label{contour}
\end{figure}

\begin{figure}[!t]
\centering
\includegraphics[width=1\linewidth, trim=0 20 0 0,clip]{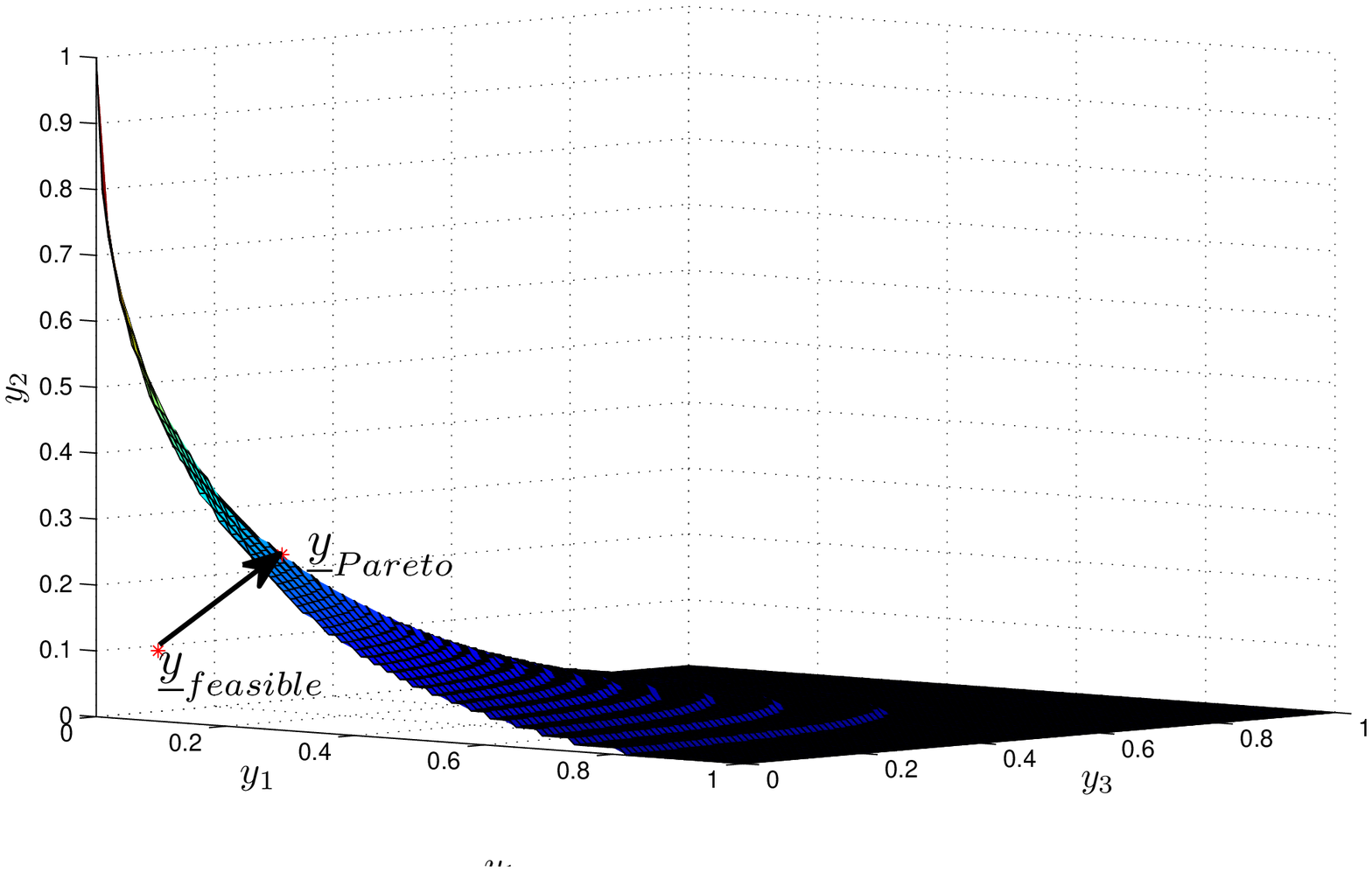}
\caption{Feasible target rate region for the 3-player chain-like topology} \label{FeasibleRegion}
\end{figure}

From Fig. \ref{contour}, it can be seen that the maximum achievable target rates in three-player chain-like topology are larger than those of the fully connected topology. For example, notice that $[y_1,y_3] = [0.15, 0.15]$ is below the target rate contour $ y_2=0.15 $ of the chain-like topology, thus the combination $[y_1,y_2,y_3] = [0.15,0.15, 0.15]$ is achievable and a stable NE can be found. However, the same combination is not achievable for the fully connected topology.

An alternative way of illustrating the feasible target rate region (the region under the mesh surface) for the three-player chain-like topology is shown in
Fig. \ref{FeasibleRegion}. The upper boundary of this feasible region is the Pareto front\cite{Pareto}, i.e., each point on the mesh surface is a target
rate combination that achieves the Pareto optimal bandwidth utilization.

\subsubsection{Improving System Sum-Rate}

Now suppose the three players have chosen $[y_1,y_2,y_3] = [0.15,0.15, 0.15]$ as their target rates. According to the previous results, they will arrive at a stable operating point $\underline{q}^*$=[0.1952, 0.2316, 0.1952].
At this operating point which is the NE, since the network is not fully loaded, intelligent players have the opportunity to further increase their throughputs until the network becomes critically stable, so as to achieve a better spectrum utilization efficiency. In the process of adjusting, all players should also ensure that the process is still governed by the underlying stability conditions defined by the generalized Aloha game.

There are many ways to achieve this and we will get different sum rates and fairness for all players. Two direct ways are: (a) each player proportionally increases its demand from $y_i$ to $ky_i$, where $k\ge 1$. (b) each player proportionally increases its transmission probability from $q_i^*$ to $bq_i^*$, where $b\ge 1$. The results are summarized in TABLE \ref{pricing}.

\begin{table*}[t]
\centering
\begin{tabular}{|c|c|c|c|c|}
\hline
• &$k_{max}$ or $b_{max}$&$\underline{y}$&$\underline{q}^*$&$\Sigma y_i$\\
\hline
original demand&1&[0.15,0.15,0.15]&[0.1952,0.2316,0.1952]&0.45\\
\hline
$\underline{y}\rightarrow k\underline{y}$&1.27&[0.1905,0.1905,0.1905]&[0.3336,0.4290,0.3336]&0.5715\\
\hline
$\underline{q}^* \rightarrow b\underline{q}^*$&1.94&[0.2086,0.1734,0.2086]&[0.3787,0.4493,0.3787]&0.5905\\
\hline
\end{tabular}
\caption{Improving sum-rate by proper pricing strategies} \label{pricing}
\end{table*}

This example shows that we can increase the sum rate of all players by proper pricing strategies or some commonly agreed target rate adjusting rules which guarantee certain criteria of fairness and maintain network stability. Conversely, the pricing strategies or target rate adjusting rules can also be used to bring the target rates of the players back to the feasible region, if the players are over demanding and the resulting network is congested.
The rationale behind this study is as follows. For future autonomous radios which compete to transmit like in the ISM band, we would like each device to equip with intelligence so that while competing to transmit, each transmission pair is also governed by the underlying rules so that maximum throughput can be achieved without affecting the network stability. This will result in a win-win situation for all transmission pairs. More vigorous design approach is currently under development.

\subsection{Spatial Reuse Gain versus Connectivity}
Consider a distributed network with $N$ players, which are randomly placed in a square region of a given area. One half of the players will have transmission range of 5 unit length, while the other half of players have transmission range of 3 unit length. We assume that all the distances between any transmitter and its designated receiver are much smaller than the distances between any two transmitters. We further assume that those players who are in each other's transmission range will have significant interference on each other, and the two nodes are said to be connected. The interference matrix can then be constructed based on the generated network topology.

\subsubsection{Performance as Player Density Increases}
\textit{Player density} is defined as the number of players per unit area. We first set $N=20$, and the player density is increased by decreasing the
spatial area under consideration. For each given player density, we run the simulation 100 times. Each time a random topology is generated, and for
simplicity, we assume that all players have the same target rate. We increase this common target rate in steps of 0.001, and run the Krasovskii's method
until the least fixed point is no longer stable. Consequently, this NE corresponds to the transmission probabilities for the players to achieve the maximum
target rate.

\begin{figure}[!t]
\centering
\includegraphics[width=0.8\linewidth]{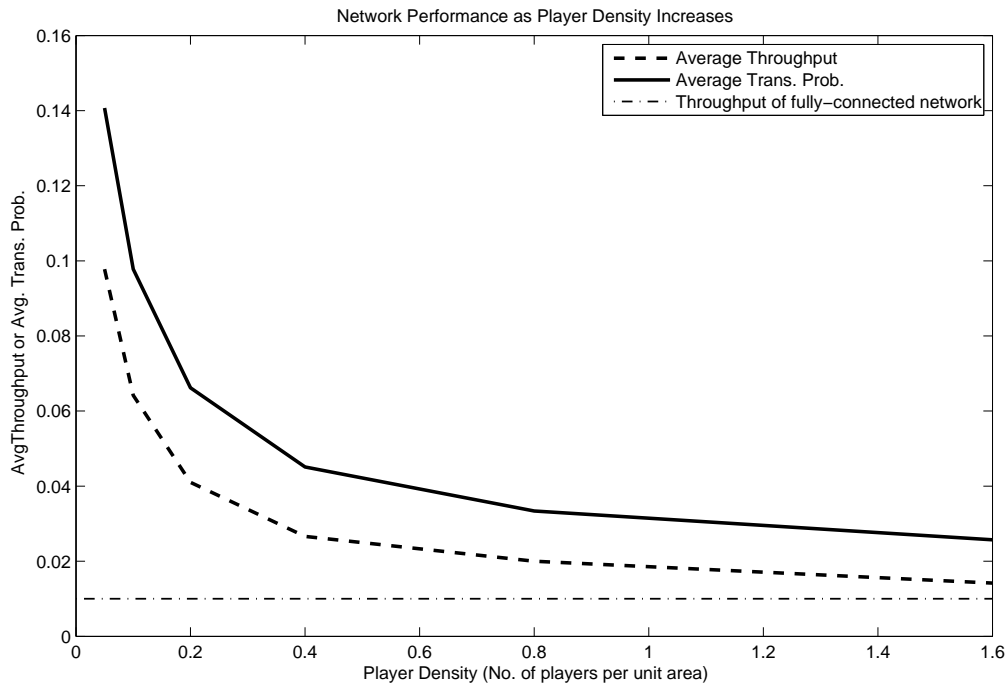}
\caption{Effect of the player density on the average throughput (20 players in decreasing spatial area)} \label{densityincrease}
\end{figure}

Fig. \ref{densityincrease} shows that both the average throughput and the average transmission probability decrease as the player density increases. The average throughput curve gradually approaches the lower limit in which all players are fully connected (which is equivalent to the conventional Aloha games).

\subsubsection{Performance as Number of Players Increases}
In this subsection we fix the player density at 0.1, and increase the number of players by increasing the spatial area.

\begin{figure}[!t]
\centering
\includegraphics[width=1\linewidth,  trim=0 0 0 0,clip]{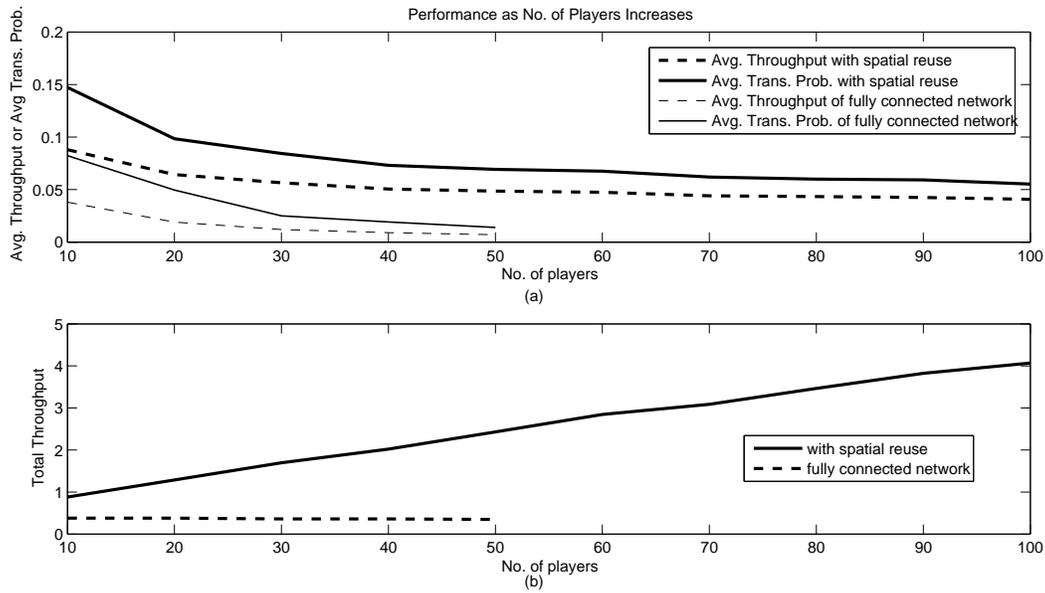}
\caption{Effect of the player density on the average throughput (Player density = 0.1, increase spatial area)} \label{numincrease}
\end{figure}

Fig. \ref{numincrease}a shows that the average throughput and transmission probability for the players decrease as the number of players increases, for both fully connected network and the generalized Aloha game.
It can be seen that the average throughput in the generalized Aloha game is significantly higher than that in a fully connected network.
The achievable average throughput of the fully connected network drops below 0.01 when there are more than 40 players. This is comparatively low when compared to the generalized Aloha game, whose average throughput stays above 0.04 even when there are 100 players. We skip the simulation for the fully connected network when the number of players is more than 50.

Fig. \ref{numincrease}b shows that the total throughput for the generalized Aloha game increases almost linearly as the number of players increases.
On the other hand, the total throughput for the fully connected network remains at a low level around 0.37.

\subsubsection{Relationship between Total Throughput and Connectivity}
Define \textit{connectivity} as the total number of links in the current network versus the total number of links in the fully connected case. In particular, connectivity equals to 1 in the conventional Aloha games. Connectivity therefore serves as an indication of spatial reuse capability. From the above observations, it can be seen that if the network is nearly fully connected, i.e., most of the players are within the interference range of each other, its throughput resembles to that of a conventional Aloha game. As the network connectivity drops, either due to decreased player density or due to a larger spatial area compared to the transmission range, the total achievable throughput increases, indicating an increased spatial reuse capability. We therefore postulate that there could exist a relationship between the reuse capability versus the network connectivity. We will use the data from the above two subsections, and present the relationship between total throughput and connectivity.

\begin{figure}[!t]
\centering
\includegraphics[width=0.8\linewidth]{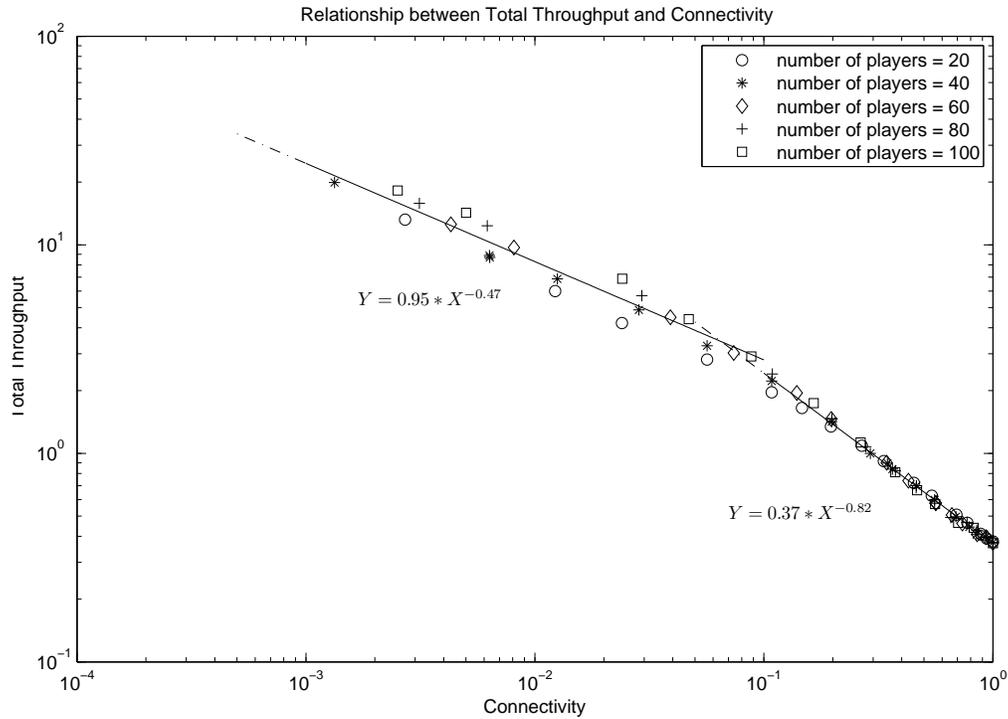}
\caption{Relationship between total throughput and connectivity} \label{TotalThrVsCon}
\end{figure}

Fig. \ref{TotalThrVsCon} shows that the total throughput decreases as the connectivity increases, regardless of the number of players involved. The relationship between total throughput (\textit{Y}) and connectivity (\textit{X}) can be approximated by an empirical formula:

\begin{equation}
Y=\left\{
    \begin{array}{l l}
      \textrm{$0.95*X^{-0.47}$} & \quad \textrm{ $0.001\le X < 0.1$}\\
      \textrm{$0.37*X^{-0.82}$} & \quad \textrm{ $0.1\le X \le 1$}\\
    \end{array} \right.
\end{equation}

Notice that when connectivity is sufficiently low (below 0.001), the network actually degenerates into several independent connected sub-networks, whose connectivity is higher than the original network. In that case, we can apply the above formula separately to each connected sub-network.

\section{Conclusions and Future Work}\label{SectionConclusion}

In this paper, we extend the slotted Aloha games to spatial reuse scenarios, namely, generalized Aloha games. We use fixed point theory and order theory to prove the existence of a unique NE in the generalized Aloha game. In particular, we use the Kleene fixed-point theorem to prove the existence of a least fixed point, which is the unique NE of the game and the most energy-efficient operating point for all players. We then propose to use the Krasovskii's method to prove the stability of the NE. After obtaining the conditions for system stability, we further prove that if the least fixed point is not stable, nor are other fixed points. These findings ensure the ease in finding the NE of a generalized Aloha game as we only need to focus on the least fixed point. If this point is stable, then all players can arrive at this NE through game iteration, by conveniently choosing $\underline{0}$ or $\underline{y}$ as the initial point.

We then show through simulation that the theory derived can be applied to large-scale distributed systems with complicated network topologies to study the maximum achievable throughput. An empirical relationship between the network connectivity and the achievable total throughput is finally obtained through simulations.

Pricing strategies or some target rate adjusting rules are required to bring the target rates within the feasible region. This paper has not yet addressed such issues for the generalized Aloha game, despite the simple illustration via the example of the three-player chain-like topology. Future work could be the design of pricing strategies or target rate adjusting rules for the players in a distributed manner to bring the target rates within the feasible region, or more desirably, toward an optimal combination of target rates which maximizes the total throughput of all players given certain fairness criteria.

\section*{Acknowledgement}
The authors would like to thank the three anonymous reviewers whose comments and suggestions have helped significantly improving the quality of the paper.

\bibliography{IEEEabrv,AlohaGames}

\end{document}